\documentclass[runningheads]{elsarticle}
\usepackage{graphicx} 
 
\usepackage[margin=1.2in]{geometry}
\usepackage{natbib}
\usepackage{hyperref}

\usepackage{appendix}
\usepackage{helvet}
\usepackage[table,xcdraw]{xcolor}
\usepackage{longtable}
\usepackage{pdfpages}

\begin{document}
\begin{frontmatter}
\title{Deep learning-based brain segmentation model performance validation with clinical radiotherapy CT}
\author[1,2]{Selena Huisman\corref{cor1}}
\ead{selena.isabelle.huisman@gmail.com}
\author[2,3]{Matteo Maspero}
\author[2]{Marielle Philippens}
\author[1,2]{Joost Verhoeff}
\author[1,2]{Szabolcs David}

\cortext[cor1]{Corresponding author}

\address[1]{Department of Radiation Oncology, Amsterdam UMC,
    De Boelelaan 1117, 1081 HV Amsterdam, The Netherlands
}
\address[2]{Department of Radiation Oncology, UMC Utrecht,
    Heidelberglaan 100, 3508 GA Utrecht, The Netherlands.
}
\address[3]{Computational Imaging Group for MR Diagnostics $\&$ Therapy, UMC Utrecht, Heidelberglaan 100, 3508 GA Utrecht, The Netherlands.
}
\begin{abstract}
    \noindent\textbf{Introduction} Manual segmentation of medical images is labor intensive and especially challenging for images with poor contrast or resolution. The presence of disease exacerbates this further, increasing the need for an automated solution. To this extent, SynthSeg is a robust deep learning model designed for automatic brain segmentation across various contrasts and resolutions. This study validates the SynthSeg robust brain segmentation model on computed tomography (CT), using a multi-center dataset. \\
    \textbf{Methods} An open access dataset of 260 paired CT and magnetic resonance imaging (MRI) from radiotherapy patients treated in 5 centers was collected. Brain segmentations from CT and MRI were obtained with SynthSeg model, a component of the Freesurfer imaging suite. These segmentations were compared and evaluated using Dice scores and Hausdorff 95 distance (HD95), treating MRI-based segmentations as the ground truth. Brain regions that failed to meet performance criteria were excluded based on automated quality control (QC) scores.\\
    \textbf{Results} Dice scores indicate a median overlap of 0.76 (IQR: 0.65-0.83). The mean volume difference is 7.79\% (CI: 6.41\%–9.18\%), with CT segmentations typically smaller than MRI-based. The median HD95 is 2.95 mm (IQR: 1.73-5.39). QC score based thresholding improves median dice by 0.1 and median HD95 by 0.05mm. Morphological differences related to sex and age, as detected by MRI, were also replicated with CT, with an approximate 17\% difference between the CT and MRI results for sex and 10\%  difference between the results for age.\\
    \textbf{Conclusion} SynthSeg can be utilized for CT-based automatic brain segmentation, but only in applications where precision is not essential. CT performance is lower than MRI based on the integrated QC scores, but low-quality segmentations can be excluded with QC-based thresholding. Additionally, performing CT-based neuroanatomical studies is encouraged, as the results show correlations in sex- and age-based analyses similar to those found with MRI.\\
\end{abstract}

\end{frontmatter}

\thispagestyle{empty}
\setcounter{page}{0}
\clearpage
\setcounter{page}{1}

\section{Introduction}

Segmentation of the human brain can be used to measure regional volumetric differences between healthy subjects and patients, aiding in visualizing and quantifying brain structures \cite{Despotović2015}. Manual segmentation requires extensive labor while providing variable performance between observers, with median Dice scores falling between 0.46 and 0.89 and median HD95 between 2.8 and 3.3 depending on the brain region \cite{Lorenzen_2021}. Consequently, automatic and reproducible segmentation models have been developed, which usually require a T1-weighted magnetic resonance imaging (MRI). However, when images are only available from computed tomography (CT), the task of tissue segmentation becomes challenging due to the inferior soft tissue contrast of CT compared to MRI. To address shortcomings, contrast-agnostic tools have been trained on artificially downsampled MRIs. Such models enable automatic and reproducible segmentation of brain images regardless of contrast and resolution, effectively enabling CT-based soft-tissue segmentation. 
Recently, SynthSeg has been proposed \cite{SynthSeg1} as a promising contrast and resolution agnostic, whole-brain segmentation model. This model, trained through exclusively synthetic images with randomized contrasts and resolutions, exemplifies this robust approach, offering satisfactory CT segmentation performance \cite{SynthSeg2}. Direct comparative validation of SynthSeg and similar models is challenging, due to the need for paired CT and MRI, in which image pair availability is limited in both healthy volunteers and patients. In radiotherapy (RT), however, patients with brain tumors routinely undergo both CT and MRI for radiation planning, making this setting ideal for validation. Usually, a large part of the patient's brain is not affected by tumors, allowing us to compare the performance of CT and MRI-based segmentation of such regions. Additionally, SynthSeg incorporates automatic quality control (QC) scores to assess the accuracy of segmented brain regions, enabling the exclusion of diseased areas from the analysis.

We have assembled a benchmark dataset from three open sources, comprising 260 paired oncological brain CT and MRI images sourced from a total of five medical centers. This diverse dataset offers a comprehensive resource for validating the SynthSeg model. While quantifying CT and MRI segmentation performance is valuable, tissue segmentations are ultimately tools in exploring various medical or scientific questions, for example the analysis of volumetric differences based on sex or age, which have been extensively explored in the human brain. \cite{brainaging,sexdifference,sexage} .

In this work, we compared the segmentation performance of SynthSeg on CT and MRI of patients with brain tumors. Additionally, we examined the volumes of various brain regions between males and females using both imaging techniques. Furthermore, we assessed how the brain volumes of patients change with age and investigated whether these results obtained from MRI segmentation are reproducible using CT-based segmentation. Although our study does not primarily investigate these well-documented differences, we employ both MRI and CT-based segmentation to replicate such findings conceptually. This approach allows us to test the viability of CT-based segmentation in neuroscientific studies, which have predominantly used MRI. This comparison aims to expand the methodologies available for such research and validate the effectiveness of CT in capturing relevant brain differences, when MRI is not available.

\clearpage

\section{Methods}

Data was sourced from three openly accessible datasets, containing patients with primary or metastatic brain tumor  for a total of 260 paired CT and MRI. 
The first dataset was the SynthRAD2023 Grand Challenge (\url{https://SynthRAD2023.grand-challenge.org/}) training dataset\footnote{Retrieved on 08-09-2023 from \url{https://zenodo.org/doi/10.5281/zenodo.7260704}.}, comprising a subset of 180 brain CT and T1w 3D gradient echo MRI and are available for public access \cite{SynthRad}. These images were collected from three Dutch hospitals: UMC Utrecht, UMC Groningen, and Radboud Nijmegen. Furthermore, we utilized similar CT-T1w paired subsets from the GLIS-RT \cite{GLIS} and Burdenko-GBM-Progression projects \cite{GBM}, from the TCIA database \cite{TCIA}, which are also available openly. The GLIS-RT data was collected at Massachusetts General Hospital, while the Burdenko data is from the Burdenko National Medical Research Center of Neurosurgery.  All CT and T1 images have been registered and rescaled to 1mm isotropic resolution, no additional processing was applied to the data. An overview of the datasets and demographics can be found in table \ref{tab:datatable}. GLIS-RT provides contrast-enhanced 3D-T1 weighed MRI and radiotherapy planning CT. Meanwhile, Burdenko-GBM-Progression includes T1-weighed MRI and topometric CT, in which the MRIs were obtained from four vendors with varying scanning protocols  and the CTs were obtained with a single scanning protocol.
In principle, our analysis could be replicated by many radiotherapy departments with access to similar data, the open nature of our sources and the mixture of contributing centers enhancing the transparency and reproducibility of our results.

Next, we used the SynthSeg tool within the Freesurfer software suite, version 7.4.1 \cite{SynthSeg2} to perform automatic brain segmentation based on 95 brain regions. The model was run in robust mode, with the integrated QC score and volume options enabled. The QC scores evaluate the reliability of each segmentation by comparing the segmented output against a learned model of what constitutes a high-quality segmentation. These scores are generated by assessing features of the segmentation against the predicted values from the 'regressor R' of the SynthSeg model, providing a numerical indication of segmentation quality. For CTs, the corresponding option was enabled to facilitate CT-specific processing. The QC scores are divided into eight categories: general white matter, general grey matter, general cerebrospinal fluid (csf), cerebellum, brainstem, thalamus, putamen \& pallidum, and hippocampus \& amygdala. The resulting data was examined using Rstudio version 2023.9.1.494 \cite{Rstudio}, applying a manual region-specific threshold for QC score exclusion instead of the static 0.65 threshold suggested by the SynthSeg manuscript \cite{SynthSeg2}. This manual thresholding approach was chosen to account for variations in QC score averages across different brain regions. In some cases the 0.65 threshold is too lenient resulting in unacceptable segmentations being included. Our manual thresholds correspond with the range of 0.55-0.75 found in the SynthSeg manuscript. Brain regions with inadequate QC scores were selectively removed from each image, preserving the remaining regions for subsequent analysis. The analysis included calculating the percentage mean volume difference for each brain region, visualized using bar plots and percentage-based Bland-Altmann plots, also known as Giavarina plots~\cite{Giavarina_2015}. Dice scores and Hausdorff 95 distances (HD95) for each brain region were computed using the segmentation\_metrics python package, based on the segmented outputs \cite{jia2024seg}. Non-overlapping regions were excluded from the analyses as they resulted in 0 Dice score and physically implausible HD95. Also, when a label was missing for a patient from either the CT or the MRI, that particular label was excluded from analyses. Note, that only CT-based segmentations resulted in any missing regions. Upon visual inspection, all missing labels in the CT or the non-overlapping regions only occurred in the tumor affected areas, for which regions SynthSeg is not meant to produce brain segmentation labels.  The segmentations were visually inspected using FSLeyes from FSL version 6.0~\cite{FSLeyes}. For visualization purposes, the brainstem, left thalamus, left hippocampus and left cerebellum white matter were highlighted in the figures.

\begin{table}[!ht]
    \centering
    \caption{An overview of the data used.}
    \includegraphics[scale=0.8]{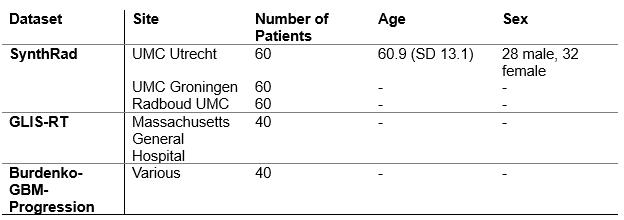}
    \label{tab:datatable}
\end{table}

\begin{figure}[!ht]
    \centering
    \includegraphics[scale=0.4]{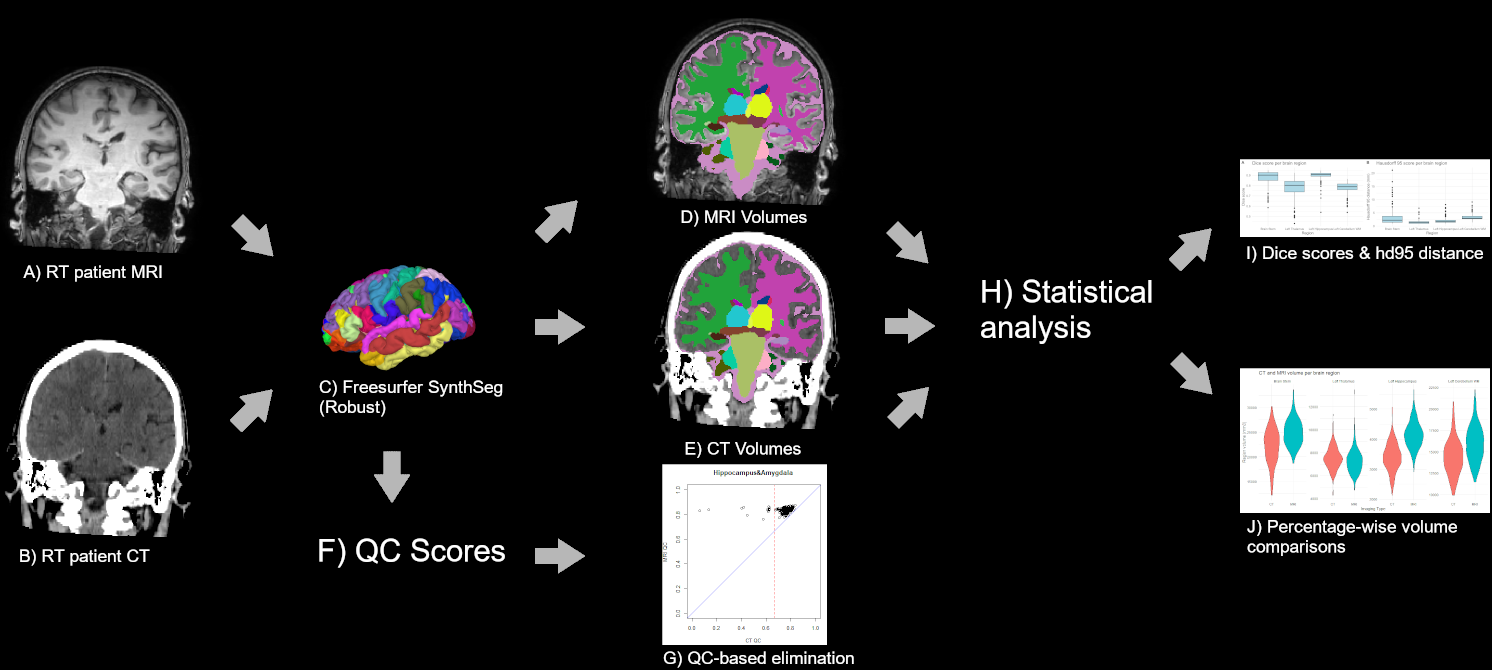}
    \caption{Pipeline of data processing and analyses. From left to right: The patient's CT and MRI are processed by SynthSeg, which outputs QC scores and segmented volumes for both image modalities. The resulting segmentations are then filtered by a QC-based threshold, after which statistical analyses is performed.}
    \label{fig:pipeline}
\end{figure}

A sex and age-based analysis was conducted using the 60 patients from UMC Utrecht. These analysis considered eight regions namely the left-thalamus, -insula cortex, -superior frontal cortex, -hippocampus, -putamen, -pallidum, -lateral ventricle, and the brainstem. The demographic details for these patients are not publicly available due to privacy regulations and were retrieved in accordance with the institutional review boards (nWMO research number 22-474). Figure \ref{fig:pipeline} provides a full overview of the analysis pipeline. 

\clearpage

\section{Results}

\begin{figure}[!ht]
    \centering
    \includegraphics[scale=0.6]{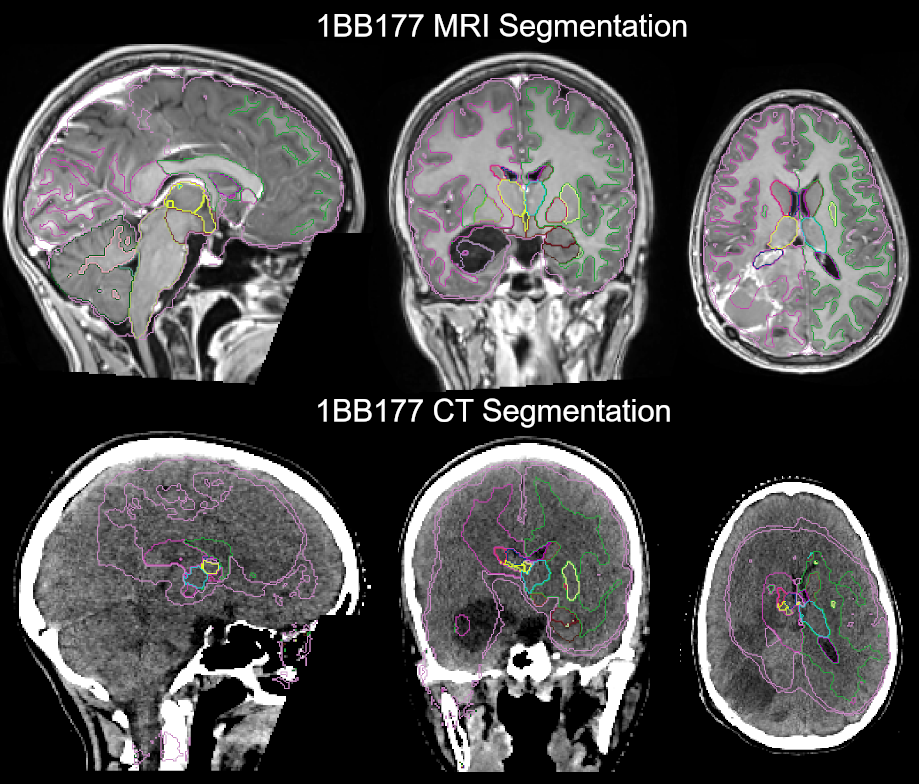}
    \caption{Segmentation of patient 1BB177 of the SynthSeg cohort. The MRI and CT segmentations are shown at the top and bottom of the image, respectively.}
    \label{fig:1bc19}
\end{figure}

\begin{figure}[!ht]
    \centering
    \includegraphics[scale=0.6]{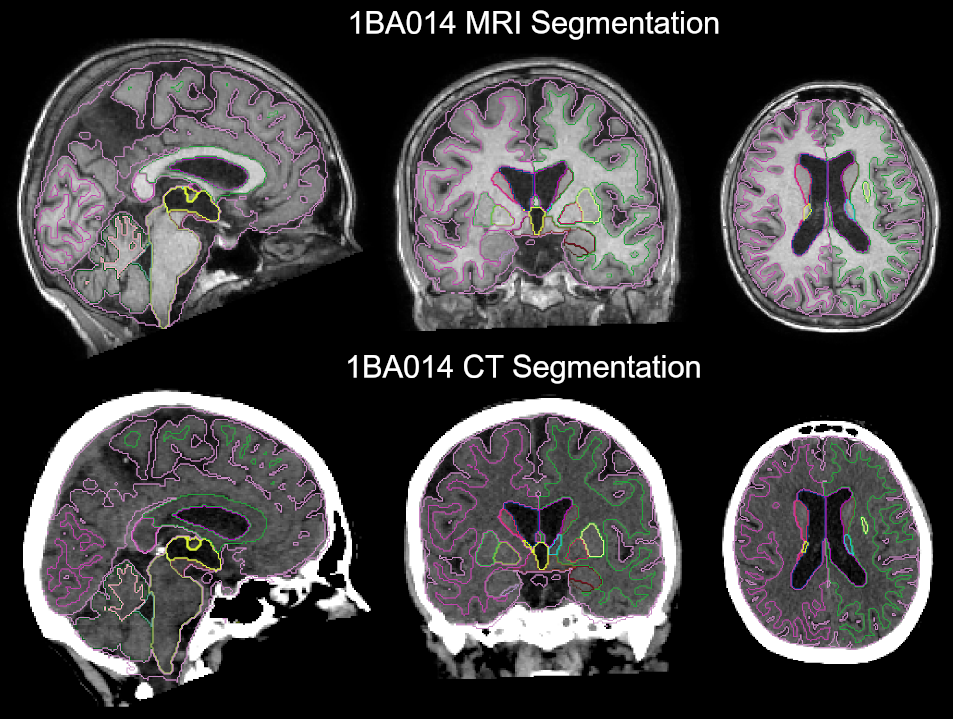}
    \caption{An example segmentation of patient 1BA014. The MRI and CT segmentations are shown at the top and bottom of the image, respectively.}
    \label{fig:1ba014}
\end{figure}

To visualize the difference between a CT  that received a low QC versus a high QC score, figure \ref{fig:1bc19} and \ref{fig:1ba014} showcase CT and MRI segmentations in two example patients. Figure \ref{fig:1bc19} shows a patient with a low mean CT QC score of 0.364 and a high MRI QC score of 0.813. While the segmentation in both image modalities fails around the tumor and treatment affected area of the brain, the CT segmentation also fails to capture accurately non-affected areas, for example the brainstem, the cerebellum and even the contralateral cortical areas relative to the tumor cavity. 

 \clearpage

Figure \ref{fig:1ba014} shows an example segmentation with high QC scores for both MRI and CT, with a mean of 0.851 and 0.839, respectively. Both segmentations capture nearly all normal-appearing parts of the brain. 

 \clearpage

\begin{figure}[!ht]
    \centering
    \includegraphics[scale=0.55]{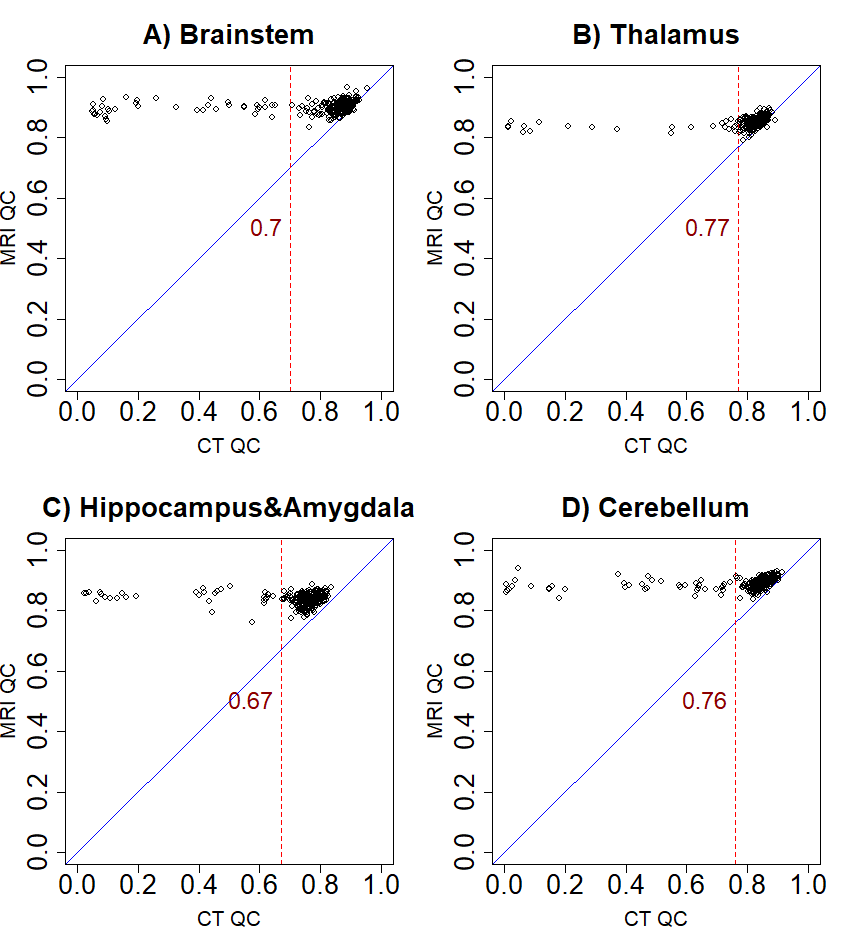}
    \caption{Four plots showing the CT QC scores against the MRI QC scores per patient. The x-axis shows the CT QC scores, while the y-axis shows the MRI QC scores. The blue line has a slope of one to indicate where the scores would be equal. The red line is the threshold selected for the exclusion of segmentations.}
    \label{fig:multiplot_threshold}
\end{figure}

MRI QC scores show low variability (SD = 0.025, 0.019 excluding CSF), for  CSF, which exhibits a higher SD of 0.072, while CT QC scores show high variability (SD = 0.139). Figure \ref{fig:multiplot_threshold} shows the comparison of CT-based and MRI-based QC scores from all 260 patients. Brain regions with CT QC scores below the corresponding threshold were subsequently excluded from the analysis, which results in a SD of 0.036. Most data points lie above the unity line (in blue) in the figures, meaning that MRI-based segmentations yield higher QC scores than those from CT-based segmentations. In all categories, MRI QC scores are significantly higher than CT QC scores. Furthermore, no brain regions were excluded based on low MRI-based QC scores if they were not already excluded due to low CT QC scores. A detailed breakdown of filtered and unfiltered mean QC scores for each brain region is available in the supplementary materials, table 1. The remaining four QC score plots can be found in supplementary materials, figure 1.

 \clearpage

\begin{figure}[!ht]
    \centering
    \includegraphics[scale=0.2]{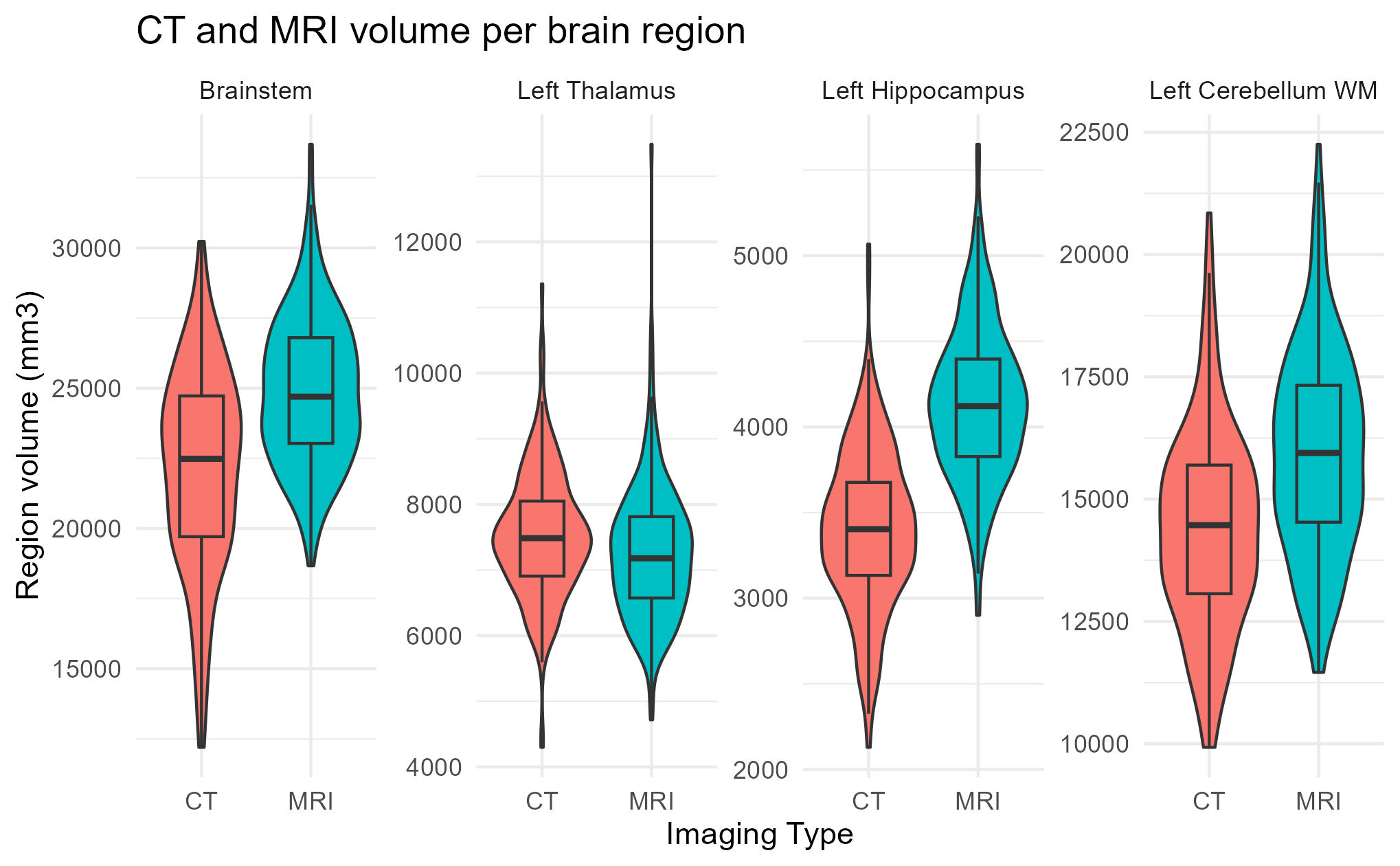}
    \caption{Four violin plots show the volume distribution of  between the MRI and CT segmentation volumes for four brain regions. On the x-axis the volume in mm3 is found, while the imaging type is listed on the y-axis. CT is colored red, while MRI is colored blue.}
    \label{fig:percentages}
\end{figure}

The distribution of volumes per brain region is shown in figure \ref{fig:percentages}. On average, the CT volumes are significantly smaller than the MRI volumes, with an average of 7.79\% (CI: 6.41\%–9.18\%) absolute difference, although the shape of the distributions is similar. 73 regions have significantly lower volume for CT compared to MRI, 16 show no statistical difference, while 6 regions (parahippocampal cortex in both hemispheres, both thalami; posteriorcingulate and paracentral in the left cortex) have a significantly larger volume for CT compared to MRI.

Detailed data on each region's numerical percentage differences and total volumes are provided in supplementary table 2. If the confidence interval of the difference between regions includes 0, it indicates no significant difference. A visualization of these percentage differences can be found in supplementary figure 1.

 \clearpage

\begin{figure}[!ht]
    \centering
    \includegraphics[scale=0.55]{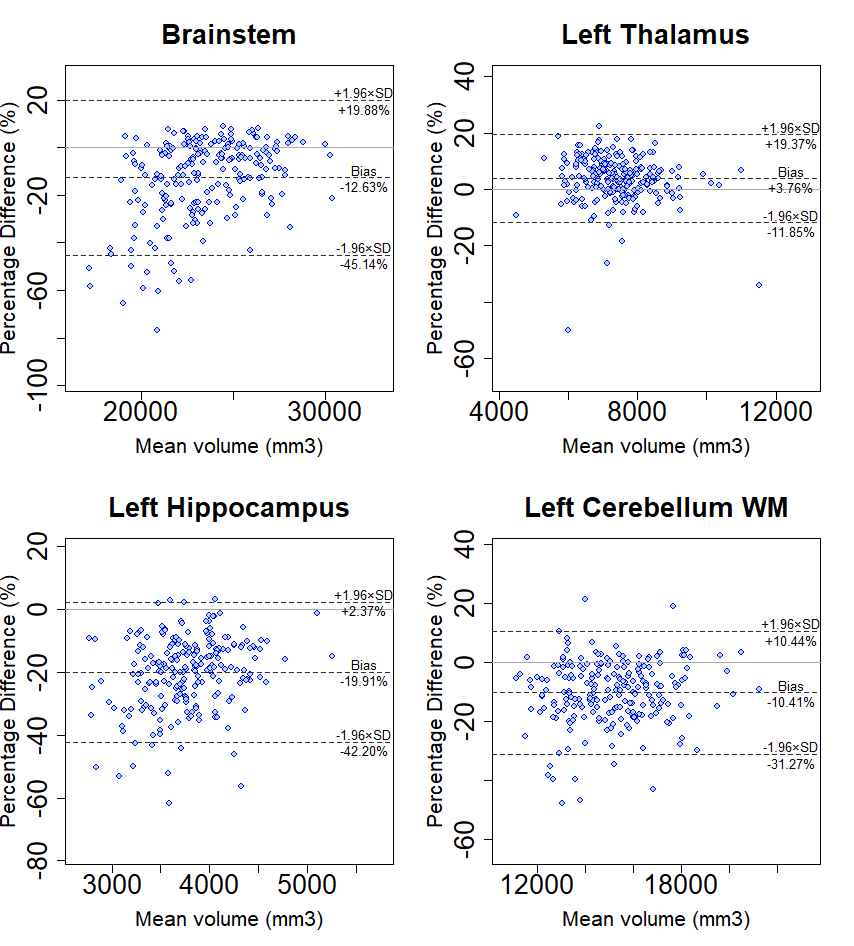}
    \caption{Four Giavarina plots for four different brain regions. The x-axis has the mean volume in mm3, while the y-axis shows the percentage difference for each segmentation. The bias and standard deviations are shown within each plot.}
    \label{fig:giabrainstem}
\end{figure}

To illustrate the bias in the distribution of the volume differences between image modalities, figure \ref{fig:giabrainstem} shows the percentage differences of volumes plotted against the mean volume for four of the 95 brain regions. The distribution tends to vary more for regions with smaller mean volumes. Additionally, CT segmentation volumes are generally smaller than those from MRI, except in the thalamus and ventricles.

\clearpage

\begin{figure}[!ht]
    \centering
    \includegraphics[scale=0.11]{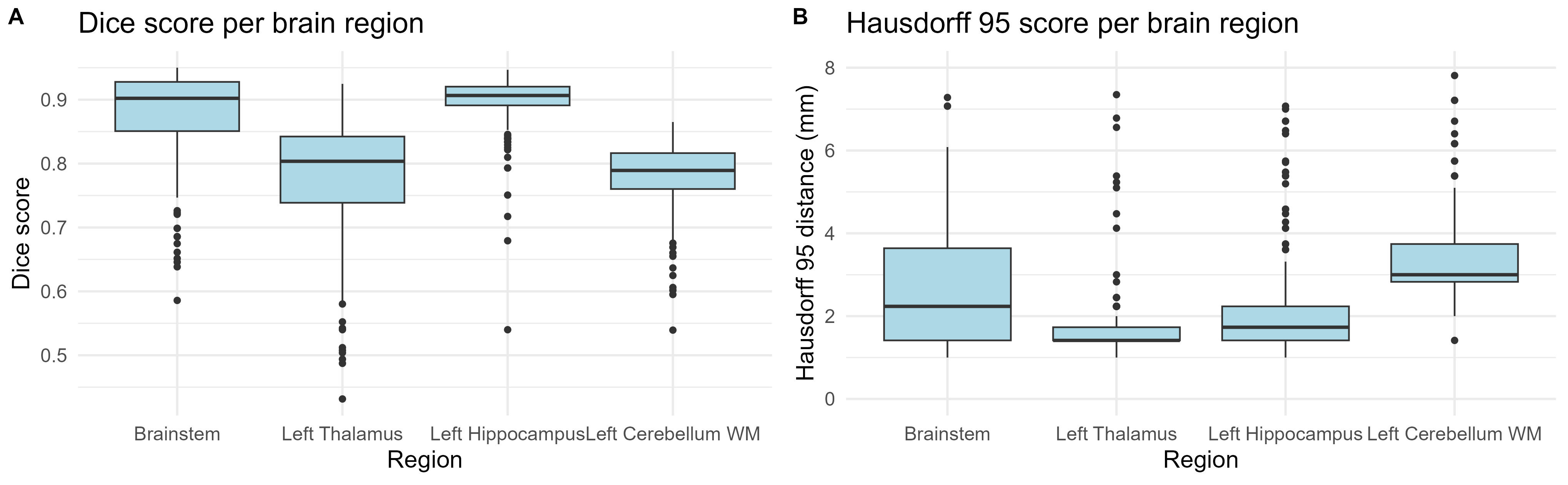}
    \caption{Two sets of boxplots show the Dice scores (A) and HD95 distance (B) for four brain regions on the y axis, respectively. The selected four brain regions are listed on the x axis. HD95 was cut off at 8mm for visual clarity, which removed some outliers for the brainstem.}
    \label{fig:dice_hist}
\end{figure}

A median Dice score of 0.76 (IQR: 0.65-0.83) was found from all regions, calculated on the overlap of each region between MRI and CT. Before QC thresholding the median overall Dice score was 0.66 (IQR: 0.50-0.78). Figure \ref{fig:dice_hist}/A shows the distribution of Dice scores for four brain regions.  White matter regions generally have higher Dice scores than regions of grey matter. Figure \ref{fig:dice_hist}/B shows the HD95 distances with a median HD95 distance of 2.95 mm (IQR: 1.73-6.16) across all regions. Before QC thresholding the median HD95 was 3. The Dice scores and HD95 distances for all regions are available in supplementary table 2.

\begin{figure}[!ht]
    \centering
    \includegraphics[scale=0.5]{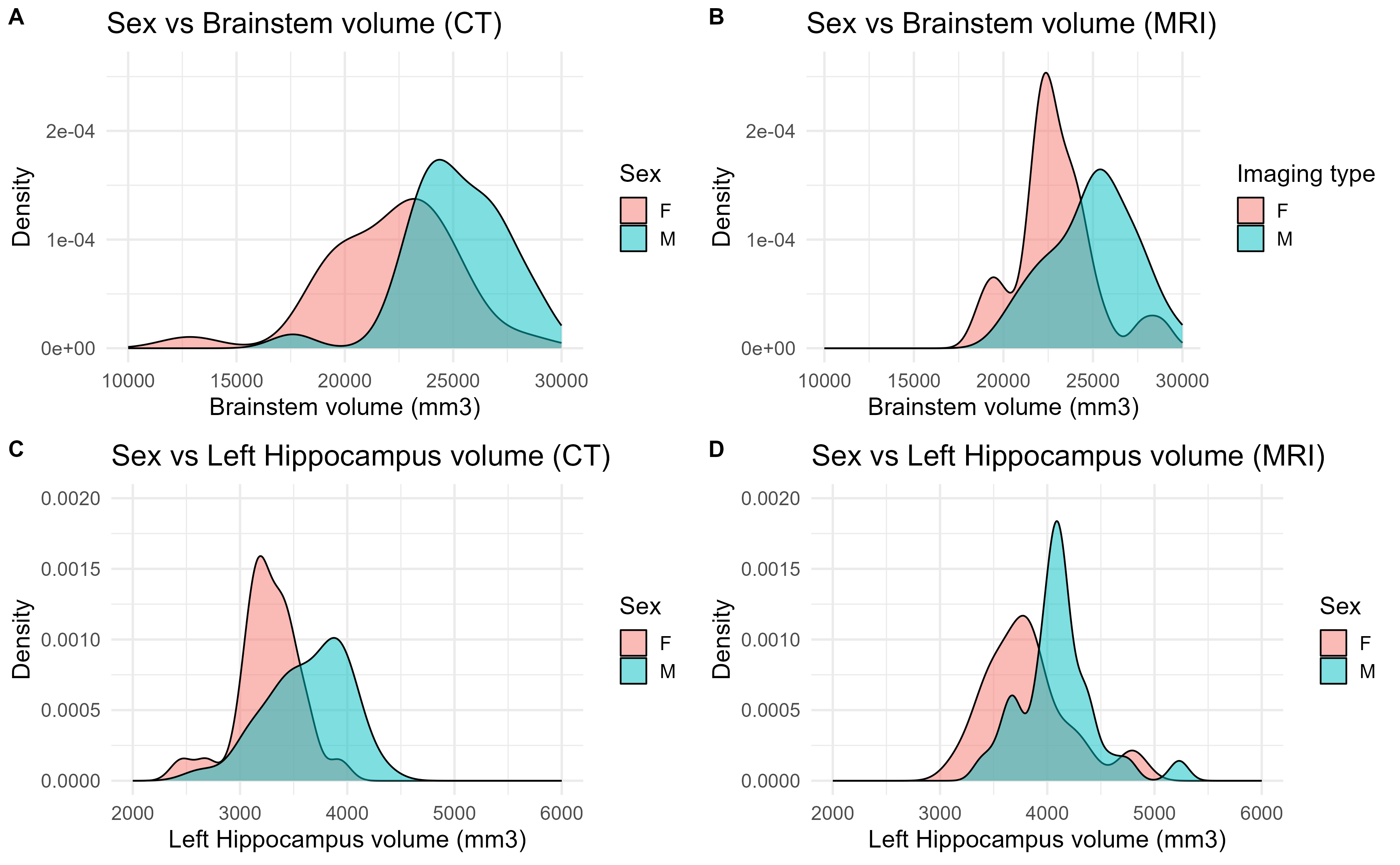}
    \caption{Four density plots comparing the volume distributions in the brainstem and hippocampus across sexes for both CT and MRI.}
    \label{fig:sexplot}
\end{figure}

After considering the effects of sex on brain region volume, density plots of the brainstem and the hippocampus volume were obtained, shown in figure \ref{fig:sexplot}. The distributions are separate for the MRI and CT volumes, while the x axes are scaled identically within the regions. The differences in the distributions between sexes established in MRI are similar to those based-on CT. The quantitative overview of the sex-based analysis is located in supplementary table 3. For all sex based analyses, the direction of the volume differences are identical with both CT and MRI based comparisons, with a mean 17\% difference overall. Of the selected regions, the brainstem shows the most considerable percentage difference of 52\% between the MRI and CT. 

 \clearpage

\begin{figure}[!ht]
    \centering
    \includegraphics[scale=0.5]{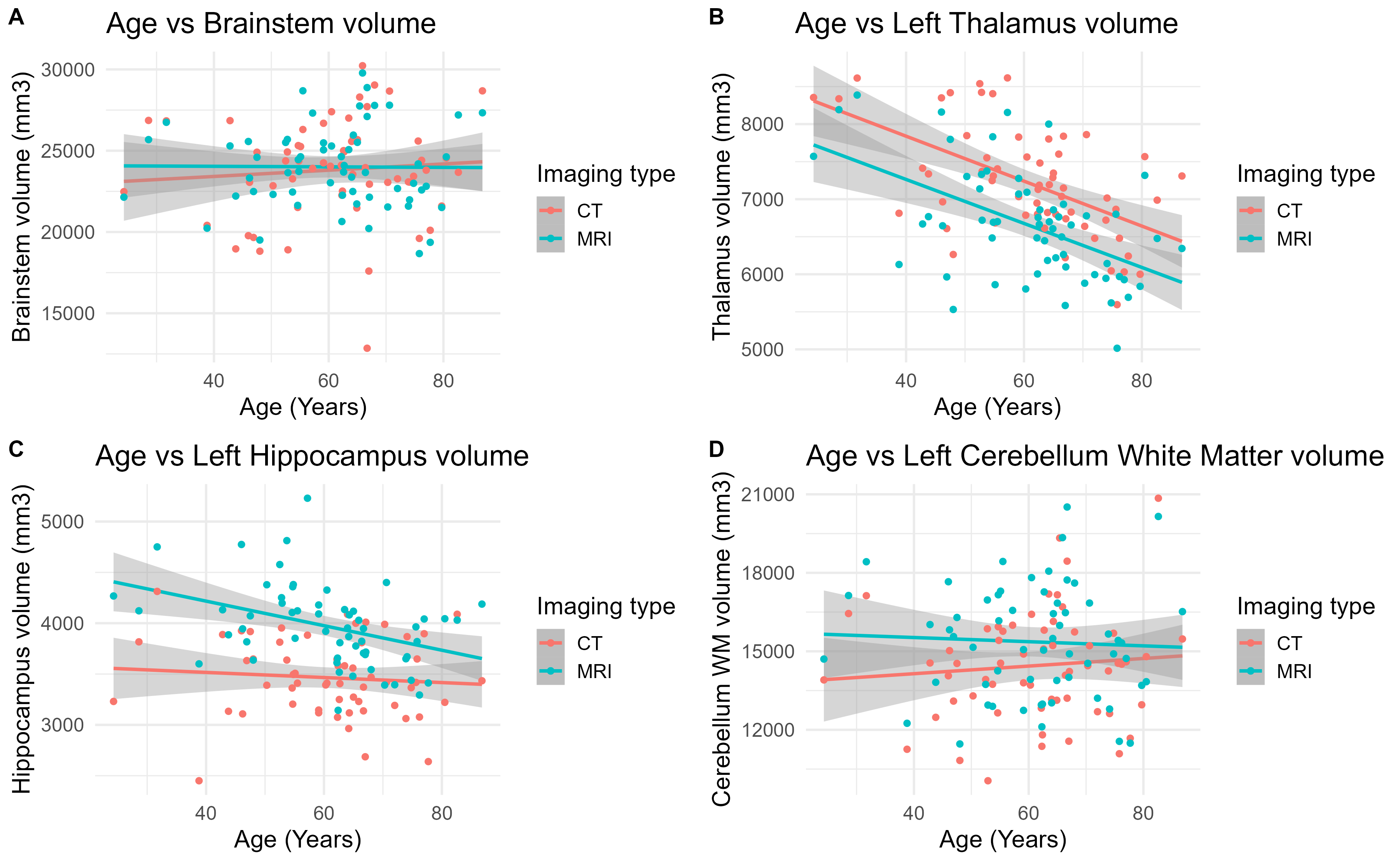}
    \caption{The figure presents four plots, each illustrating the relationship between brain region volume and age in four different brain regions. Orange points and lines denote CT data, whereas blue points and lines represent MRI data. The points mark the volumes from individual segmentation, and the lines indicate the linear regression trajectories for each imaging modality, complete with their respective confidence intervals.}
    \label{fig:ageplot}
\end{figure}

The results for the age-based analysis are shown in figure \ref{fig:ageplot} . Similarly to the sex-based analysis, most of the MR-based results and relationships are preserved in the CT-based analyses. In four selected regions, the volume of the brain areas are plotted against patient age. A linear regression slope with confidence intervals is used to visualize the effect of aging on brain region volumes, separately for CT and MRI-segmented volumes. All regions, except the brainstem, show the same direction in the volume change-age relationship. For these regions, the mean absolute difference between the regression slopes is 10\%. The brainstem show non-significant regressions in both CTs and MRIs with different slope directions. However, these differences are negligible when adjusted for the brainstem's size. The quantitative summary of the age-based analysis is also located in supplementary table 3.

 \clearpage

\section{Discussion}

In this work, we conducted a comprehensive analysis to evaluate the performance of the SynthSeg model for CT brain segmentations against MRI-based ones. The quantitative analysis suggests that SynthSeg delivers CT segmentation performance similar to the range of inter observer variability in manual segmentation, particularly in scenarios with minimal anomalies or data censoring. Should segmentation failure occur for any reason, it is reflected in the QC score, allowing for the rejection of unreliable segmentations. Nonetheless, CT segmentations typically differ from MRI-based baseline, as evidenced by the Dice score. Generally, CT QC scores are lower than those for MRI, implying superior performance in MRI segmentations. The difference is marginal in certain brain regions, such as the brainstem or the thalamus (0.037 and 0.022, respectively). Still, the disparity is more pronounced in others, such as general grey and white matter, with mean filtered QC score differences of 0.098 and 0.077, respectively. However, these regions occupy large portion of the brain and are nearly always affected by lesions. This observation aligns with the expectation of lower soft tissue contrast in CT.

The application of QC scores significantly improved the Dice score and HD95 metrics by filtering out regions unsuitable for this analysis. Typically, these are areas with abnormal tissue, such as those affected by tumors or surgical cavities. Additionally, regions like the lateral ventricles and grey matter exhibit highly variable Dice scores and regions that do not overlap despite adequate QC scores. This may suggest that QC scores are not always reliable indicators of segmentation quality in some brain regions, though they are generally effective. However, the HD95 distance indicates that segmented regions are always relatively close to each other, unless a certain region is not segmented at all. Correlations between demographic factors such as sex or age and brain volume remain consistent in MRI and CT segmentations, with a 13\% mean absolute difference for age-based analyses. This suggests that SynthSeg's robustness could improve the applicability of CT databases for research. The ability to reliably segment CT comparably to MRIs opens the door to revisiting archived clinical datasets.

Our study, while comprehensive, has certain limitations. It primarily utilizes 260 paired images from radiotherapy patients. Despite this, the ability to exclude diseased and censored brain regions using QC scores suggests that our findings could have broader applicability in settings requiring brain region segmentation. It is anticipated that segmentation performance would be enhanced in the absence of diseased brain areas. One limitation regarding using MRI as a basis for comparison is that MRI contain distortions, which are not present in CT. This could cause a discrepancy in the Dice scores and HD95 distances between MRI and CT. Keeping in mind that current investigations did not correct for MRI distortions, our expectation is that if we correct for MRI distortions beforehand, segmentation performance would improve and HD95 and Dice would be lower. However, since the Dice scores and HD95 still fall within a similar range as the interobserver variation, we believe that CT performance will still be similar to the interobserver variability with these distortions taken into account.

This work confirms that SynthSeg provides a valuable tool for utilizing CT data in research applications. Additionally, SynthSeg and related tools could be used for adaptive RT applications, possibly supporting these treatment methods within RT, where MRI is not available for soft tissue contrast \cite{Tseng2024,Matsuyama2022,Guevara2023}. Due to the multi-center, international nature of the data used in the study, the results should be generalizable to other datasets. Except for the sex and age for the UMC Utrecht cohort, the rest of the data is publicly available, allowing for full transparency over the majority of the analysis. For future work, validating SynthSeg's robustness with lower quality CT, such as cone-beam CT, could determine if the model can effectively extract data from a wide-range of applications and even archived datasets. Additionally, the model's applicability could be further generalized by acquiring paired CT/MR data from healthy individuals. Our findings suggest that SynthSeg can be effectively utilized to obtain segmentation data from CT, enhancing research capabilities.

\section{Conclusion}

The robust version of SynthSeg demonstrates segmentation of brain CT performance comparable to ranges found in interobserver variation. While CT-based segmentation typically underperforms compared to MRI-based segmentation, the model's automated provision of QC scores is a valuable feature. It allows for the exclusion of regions with substandard quality, whether due to low image quality or abnormal tissue, such as tumors. The relationships between sex, age, and brain region volumes are preserved when performed using CT, suggesting that the segmentations offer comparable insights. Therefore, SynthSeg could be effectively employed for segmenting CTs in scenarios where absolute precision is not a critical requirement.

\clearpage
\bibliography{bibliography.bib}

\end{document}